\documentstyle[12pt]{article}
\newtheorem{theorem}{Theorem}
\newtheorem{itlemma}{Lemma}[section]
\newtheorem{itproposition}[itlemma]{Proposition}
\newtheorem{itcorollary}[itlemma]{Corollary}
\newtheorem{itremark}[itlemma]{Remark}
\newtheorem{itremarks}[itlemma]{Remarks}
\newtheorem{itdefinition}[itlemma]{Definition}
\newtheorem{itexample}[itlemma]{Example}

\newenvironment{lemma}{\begin{itlemma}\rm}{\end{itlemma}} 
\newenvironment{remark}{\begin{itremark}\rm}{\end{itremark}} 
\newenvironment{remarks}{\begin{itremarks} \rm}{\end{itremarks}}
\newenvironment{corollary}{\begin{itcorollary}\rm}{\end{itcorollary}}
\newenvironment{proposition}{\begin{itproposition}\rm}{\end{itproposition}}
\newenvironment{definition}{\begin{itdefinition}\rm}{\end{itdefinition}}
\newenvironment{example}{\begin{itexample}\rm}{\end{itexample}}
\newenvironment{fact}{\noindent {\em Fact}. \ \ }{\hfill \medskip}
\newenvironment{proof}{\noindent {\em Proof}.\ \
}{\hspace*{\fill}$\Box$\medskip}
\newenvironment{claim}{\noindent {\em Claim}. \ \ }{\hfill \medskip}
\newcommand{\be}[1]{\begin{equation}\label{#1}}
\newcommand{\ee}{\end{equation}}
\newcommand{\bl}[1]{\begin{lemma}\label{#1}}
\newcommand{\br}[1]{\begin{remark}\label{#1}}
\newcommand{\brs}[1]{\begin{remarks}\label{#1}}
\newcommand{\bt}[1]{\begin{theorem}\label{#1}}
\newcommand{\bd}[1]{\begin{definition}\label{#1}}
\newcommand{\bp}[1]{\begin{proposition}\label{#1}}
\newcommand{\bc}[1]{\begin{corollary}\label{#1}}
\newcommand{\bfact}[1]{\begin{fact}\label{#1}}
\newcommand{\bex}[1]{\begin{example}\label{#1}}
\newcommand{\ec}{\end{corollary}}
\newcommand{\efact}{\end{fact}}
\newcommand{\eex}{\end{example}}
\newcommand{\el}{\end{lemma}}
\newcommand{\er}{\end{remark}}
\newcommand{\ers}{\end{remarks}}
\newcommand{\et}{\end{theorem}}
\newcommand{\ed}{\end{definition}}
\newcommand{\ep}{\end{proposition}}
\newcommand{\epr}{\end{proof}}
\newcommand{\bpr}{\begin{proof}}
\newcommand{\bcl}{\begin{claim}}
\newcommand{\ecl}{\end{claim}}

\newcommand{\bi}{\begin{itemize}}
\newcommand{\ei}{\end{itemize}}
\newcommand{\ben}{\begin{enumerate}}
\newcommand{\een}{\end{enumerate}}
\newcommand{\text}[1]{\hbox{\rm \ #1\ \/}}

\newcommand{\vs}{\vspace{0.25cm}}

\begin{document}

\begin{center}
{\Large {Controllability, Observability 
and Parameter Identification of
two coupled spin 1's\\ }}
\end{center}

\bigskip

\begin{center}

{Domenico D'Alessandro \\ \vs Department of Mathematics\\ Iowa
State University \\ Ames, IA 50011,  USA\\ Tel. (+1) 515 294
8130\\ email: daless@iastate.edu}

\end{center}

\begin{abstract}

In this paper,  we study the control theoretic properties of a couple of
interacting spin $1$'s driven by an electro-magnetic field. 
In particular,  we assume  that it is possible to observe 
 the expectation value of the total magnetization and we study
controllability, observability and parameter identification of  these
systems. We give conditions for  controllability and  observability  and
characterize the classes of equivalent  models which have  the same
input-output behavior. The analysis is motivated
by the  recent interest in three level systems 
in quantum information theory and quantum 
cryptography as well as by the problem of modeling
molecular magnets as spin networks.  

\end{abstract}

\section{Introduction}
\label{intro}


In recent years, there have been several proposals to use three level
systems, the so-called {\it qutrits}, in quantum information theory. The
proposals concern the use of these systems  as
building blocks for  protocols in quantum cryptography \cite{Durt} and
communication \cite{Brukner} as well as for the encoding of two logic
qubits \cite{Grudka}. They also have been used to study fundamental
questions in quantum mechanics such as { entanglement measures}
 \cite{Lorenza}, \cite{Cereceda}, \cite{BinFu}. A study of control of
three level systems was considered in \cite{Ugo}. From a quantum 
control perspective, a system of two coupled 
three level systems represents the next more
difficult case after the well studied system of coupled spin
$\frac{1}{2}$'s \cite{ioMCSS}, \cite{Khaneja}, \cite{Rama}. 
Motivation  to study
these  systems  also comes from the problem of modeling {\it molecular
magnets}. These novel materials \cite{Awsh}, \cite{Barbara}, \cite{Fort},
\cite{luban}, \cite{Meyer} are of interest in many applications as
nanosize magnets as well as for fundamental studies in quantum mechanics
and biology. They are modeled as networks of interacting spins. Spin
$1$'s are a very common example of three level systems. 
Examples are the nuclear spins of the 
naturally occurring  isotopes $^6Li$, $^2H$,  $^{14}N$.

\vs

We shall study  the
control-related properties,  namely {\it controllability}, {\it
observability} and {\it parameter identifiability}, 
for a pair
of interacting spin $1$'s particles. To be more specific, we shall
consider  an Heisenberg spin model with Hamiltonian given by    
\be{dinamica}
H(t):= i(A+B_xu_x(t)+B_yu_y(t)+B_zu_z(t)), \ee with \be{dinamica1}
\begin{array}{ccl}
A &:= &  -iJ_{12}(\sum_{j=x,y,z} 
\bar \sigma_j \otimes \bar \sigma_j), \\
  & & \\
B_{v}&:= & (\gamma_1 \bar \sigma_v \otimes {\bf 1}+ \gamma_2 {\bf 1}
\otimes \bar {\sigma_v}), \ \ \text{ for }
v=x,\, y, \text{ or } z. \end{array} \ee
Here $J_{12}$ is the exchange constant, $\gamma_1$ and $\gamma_2$ are
the gyromagnetic ratios of particle $1$ and $2$,
respectively; $u_{x,y,z}$ are the ${x,y,z}$ time-varying components of
the input electro-magnetic field;  $\bf 1$ is the $3 \times 3$ identity 
matrix. $\bar \sigma_{x,y,z}$ are the spin matrices spanning the three
dimensional representation of $su(2)$ \cite{sakurai} 
\be{S7}
\bar \sigma_x
=i\frac{1}{2}\pmatrix{0 & \sqrt{2} &0 \cr 
\sqrt{2} & 0 & \sqrt{2} \cr
0 & \sqrt{2} & 0}, 
\ee
\be{S8}
\bar \sigma_y
=i\frac{1}{2}\pmatrix{0 &- \sqrt{2}i  &0 \cr 
\sqrt{2}i & 0 & -\sqrt{2}i \cr
0 & \sqrt{2}i & 0}, 
\ee
\be{S9}
\bar \sigma_z
=-i\pmatrix{1 & 0 & 0\cr 0 & 0 & 0 \cr 0 & 0 & -1}. 
\ee
The total magnetization for the state
$\rho$ in the direction $v=x,y,z$ is
 given by 
\be{tot}
M_v=Tr(S_v^{TOT}\rho), 
\ee
where $S_v^{TOT}=\bar \sigma_v \otimes {\bf 1}+ 
{\bf 1} \otimes \bar \sigma_v$,
$v=x,y,z$. Recall that the density matrix (state) of the system 
$\rho$ satisfies the Liouville's equation \cite{sakurai}
\be{Liouville}
\dot \rho=-i [H(t),\rho].  
\ee
We are interested in the Heisenberg Hamiltonian (\ref{dinamica}) because
we have in mind applications to 
spin Hamiltonians modeling the dynamics of molecular
magnets \cite{Awsh}, \cite{Barbara}, \cite{Fort},
\cite{luban}, \cite{Meyer}. 
However,  the methods presented in this paper  can be generalized to different
types of coupled three level systems as for example two spins $1$'s with
interaction different from the one modeled in (\ref{dinamica1}) or
cases where one component of the magnetic field is held constant. The
main tool is a Cartan decomposition of the Lie algebra $su(3)$,
described in Section 2,  which gives a decomposition of  higher
dimensional Lie algebras constructed with tensor products of matrices in
$su(3)$. We begin by stating the definitions concerning controllability,
observability and parameter identification with reference to the system
we want to study. 

\vs

\bd{Controllabilita} An $n-$level quantum system 
is controllable if it is possible to drive 
the evolution operator to any value in the special unitary group
$SU(n)$.  
\ed
Controllability can be checked \cite{Tarn} by verifying that the Lie
Algebra generated by the matrices defining the dynamics ($A,B_x,B_y,B_z$
in (\ref{dinamica}), (\ref{dinamica1})) contains $su(n)$ (in this case
$su(9)$). 
\bd{Osservabilita} Denote by $\rho(t, \vec u, \rho_0)$ the trajectory
corresponding to an initial state $\rho_0$ and control(s) $\vec u$. Let
$S$ be the matrix corresponding to the output of the system 
$Tr(S\rho(t, \vec u, \rho_0))$ (in our case $S=S_v^{TOT}$, $v=x,y,z$ in
(\ref{tot})). Then the system is observable if 
$Tr(S\rho(t, \vec u, \rho_0))=Tr(S\rho(t, \vec u, \rho_0'))$,  for every
$t$ and control $\vec u$, implies $\rho_0=\rho_0'$. 
\ed
This definition of observability refers to identification  
of the initial state by a measure of the expectation value of an
observable. Observability for quantum mechanical systems in these terms 
was studied in \cite{CDC2003}. If $\cal L$ is the dynamical Lie Algebra
(generated by $A$, $B_{x,y,z}$ above for our system), an $n-$level  
system with
output $S$ (assumed w.l.g. traceless) ($S_v^{TOT}$ in our case) 
is observable if and only if 
\be{condob}
{\cal V}:=\oplus_{k=0}^\infty ad_{\cal L}iS=su(n). 
\ee 
Here and in the following 
$\oplus$  denotes the sum of vector spaces (not necessarily
direct sum). In the controllable case $su(n) \subset {\cal L}$ 
and (\ref{condob}) is verified. 

\vs

Now consider two models $\Sigma$ and $\Sigma'$ of Heisenberg
spin $1$'s. These models may  differ by  the parameters $J_{12}$ and
$\gamma_{1,2}$. They may  also have different initial states say
$\rho_0$ and $\rho_0'$. Therefore we often consider the pair 
$(\Sigma, \rho_0)$ and the pair $(\Sigma', \rho_0')$. We investigate
whether it is possible to distinguish state and parameters by an
experiment involving control with an input field and a measurement of
the output $S_v^{TOT}$. This problem is motivated by recent results on
the isospectrality of Heisenberg Hamiltonians which 
showed the impossibility  to  
distinguish the parameters in the
Hamiltonian by a measure of thermodynamic properties 
\cite{Luban}.    We call 
$\rho(t):=\rho(t, u_x,u_y,u_z, \rho_0)$ a general trajectory for 
$(\Sigma, \rho_0)$ and $\rho'(t):=\rho(t, u_x,u_y,u_z, \rho_0')$ the
corresponding trajectory (with the same control) for $(\Sigma',
\rho_0')$. We give the following definition \cite{Spin12}. 
\bd{equivalenti}
Two pairs $(\Sigma, \rho_0)$ and $(\Sigma', \rho_0')$ are equivalent if 
\be{eqio}
Tr(S_v^{TOT}\rho(t))=Tr(S_v^{TOT}\rho'(t)), 
\ee 
for every trajectory $\rho$ and corresponding (with the same control) 
trajectory $\rho'$. 
\ed
The question of whether or not it is possible to distinguish two models
using a reading of the total magnetization will be posed by describing
the classes of equivalent pairs model-initial state. If $\rho_0$ and
$\rho_0'$ are scalar matrices so are $\rho(t)$ and $\rho'(t)$ for every
$t$ so the outputs (\ref{eqio}) are identically zero independently of
the model. We shall exclude this degenerate case in the following
treatment.

\vs

The rest of the paper is organized as follows. 
In the next section,  we describe some
properties and a decomposition of the Lie algebra $su(3)$ that will be
used in the following. The question of controllability and observability
 is tackled in
Section 3, where we prove that the system is controllable and observable
if and only if $\gamma_1 \not= \gamma_2$ and $J_{12} \not=0$. In Section
4 we give a description of the classes of equivalent pairs  which,  we
prove,  consist of only two elements. Some concluding remarks 
 are presented  in Section 5.   

\section{Properties of the Lie algebra $su(3)$}
\label{prop}
The Lie algebra $su(3)$ appears in several areas of quantum physics. As
a result, it has been  extensively studied in the mathematical physics
literature (see e.g. \cite{cahn}, \cite{georgi}). We describe here some
properties that are important for our treatment. 
We consider a canonical, orthogonal 
basis of $su(3)$ given by the three
matrices 
\be{Com1}
\sigma_x=\pmatrix{0 & i & 0\cr i & 0 & i \cr 0 & i & 0}, \qquad 
\sigma_y=\pmatrix{0 & 1 & 0\cr -1 & 0 & 1 \cr 0 & -1 & 0}, \qquad \sigma_z=\pmatrix{-i & 0 & 0\cr 0 & 0 & 0 \cr 0 & 0 & i}, 
\ee
(compare with (\ref{S7})-(\ref{S9})), and the
matrices 
\be{Co4}
R:=\pmatrix{0 & 0 & i\cr 0 & 0 & 0\cr i& 0 & 0}, \qquad 
Q:=\pmatrix{0 & 0 & 1 \cr 0 & 0 & 0 \cr -1 & 0 & 0}, \qquad 
T:=\pmatrix{i & 0 & 0\cr 0 & -2i & 0 \cr 0 & 0 & i}, 
\ee
\be{Co7}
V:=\pmatrix{0 & 1 & 0 \cr -1 & 0 & -1 \cr 0 & 1 & 0}, \qquad 
U:=\pmatrix{0 & i & 0 \cr i & 0 & -i \cr 0 & -i & 0}.  
\ee
The matrices $\sigma_{x,y,z}$ span a Lie algebra 
isomorphic to  $su(2)$ which we denote by $\cal S$. In particular we
have 
\be{hj}
[\sigma_x,\sigma_y]=2 \sigma_z, \quad [\sigma_y,\sigma_z]=\sigma_x,
\quad [\sigma_z,\sigma_x]=\sigma_y. 
\ee 
The matrices
$R,Q,T,V,U$ along with multiples of the $3 \times 3$ identity $\bf 1$ 
span the
orthogonal complement of $\cal S$ in $u(3)$, which we denote by 
${\cal S}^{\perp}$. The following tables summarize the remaining commutation
relation for $su(3)$ in terms of the basis elements we have defined 
\be{Tavolacom}
\begin{tabular}{|l|l|l|l|l|} \hline 
$[,]$ &R & Q  & T  & V  \\ \hline 
  Q &$-2 \sigma_z$ &   &   &    \\ \hline
 T &0 & 0  &   &    \\ \hline
 V &$\sigma_x$ & $\sigma_y $  &$ 3 \sigma_x$  &    \\ \hline
 U &$ \sigma_y$  & $-\sigma_x$   & $-3 \sigma_y$   & $2 \sigma_z$    \\ \hline
\end{tabular} \qquad 
\begin{tabular}{|l|l|l|l|l|l|} \hline 
$[,]$ &R & Q  & T  & V & U  \\ \hline 
  $\sigma_x$  & $-V$ & $U$  & $-3V$   & $2T+2R$  & $-2Q$     \\ \hline
 $\sigma_y$ & $-U$ & $-V$  & $3U$  & $2Q$  & $-2T+2R$     \\ \hline
 $\sigma_z$ &$2Q$   & $-2R$   &$ 0$   & $-U$ & $V$     \\ \hline
\end{tabular} 
\ee
From these tables, it follows 
\be{Co9}
[{\cal S}, {\cal S}] = {\cal S}, \quad 
[{\cal S}^\perp , {\cal S}] = {\cal S}^\perp /{span \{i {\bf 1} \}}  \quad 
[{\cal S}^\perp , {\cal S}^\perp ] = {\cal S}. 
\ee
We also have that for any matrix in the set  $\{R,Q,T,V,U \}$, say $L$ 
\be{plplp}
\bigoplus_{k=0}^\infty ad_{\cal S} L= {\cal S}^\perp /{span \{i {\bf 1} \}}. 
\ee

The anticommutation relations are summarized in the following tables 
\be{Tavolaanticom}
\begin{tabular}{|l|l|l|l|l|l|} \hline 
$-i \{, \}$ & R & Q  & T  & V & U  \\ \hline 
R  & $\frac{4}{3}i{\bf 1}+\frac{2}{3} T$ &   &   &  &  \\ \hline 
Q &$0 $ & $\frac{4}{3}i{\bf 1}+\frac{2}{3} T$    &   &  &  \\ \hline
 T & $2R$ & $2Q$  &$4 i {\bf 1} -2T$   &  &   \\ \hline
 V &$V$  & $-U $  &$ -V $  &$\frac{8}{3}i{\bf 1}- \frac{2}{3} T +2R$  &  \\ \hline
 U &$ -U $  & $-V$   & $-U$   & $-2Q$ & $\frac{8}{3}i{\bf
1}-\frac{2}{3}T -2R$
    \\ \hline
\end{tabular} \qquad 
\ee
$$
\begin{tabular}{|l|l|l|l|l|l|} \hline 
$-i\{,\}$ & R & Q  & T  & V & U  \\ \hline 
  $\sigma_x$  & $\sigma_x$ & $\sigma_y$   & $-\sigma_x$   
& 0 & $2 \sigma_z$      \\ \hline
 $\sigma_y$ & $-\sigma_y$  & $\sigma_x$   & $-\sigma_y$  & $2 \sigma_z$

& $0$      \\ \hline
 $\sigma_z$ & $0$    & $0$   & $2 \sigma_z$   & $\sigma_y$ & $\sigma_x$
    \\ \hline
\end{tabular} \qquad 
\begin{tabular}{|l|l|l|l|} \hline 
$-i\{,\}$ & $\sigma_x$  & $\sigma_y$   & $\sigma_z$  \\ \hline 
  $\sigma_x$  & $\frac{8}{3}i{\bf 1}- \frac{2}{3}T+2R$ &    & \\ \hline
 $\sigma_y$ & $2Q$  & $\frac{8}{3}i{\bf 1}- \frac{2}{3}T- 2R$   &   \\ \hline
 $\sigma_z$ & $U$    & $V$   & $\frac{4}{3}i{\bf 1}+\frac{2}{3}T$  \\\hline 
\end{tabular}. 
$$
We have 
\be{nice}
i\{{\cal S},{\cal S}\}={\cal S}^\perp, \quad 
i\{{\cal S}^\perp , {\cal S}\} = {\cal S} \quad
i\{{\cal S}^\perp , {\cal S}^\perp \} = {\cal S}^\perp.  
\ee
In the following, we  denote by $\sigma$ a generic 
element of $i {\cal S}$ and by $S$ a generic element 
of $i {\cal S}^\perp $. Therefore, $\sigma$ and $S$ are Hermitian
matrices. The decomposition of $u(3)$ which we have introduced in this
section has consequences for decompositions of higher dimensional
spaces. We shall use this in the following sections, in particular in
Section \ref{ParId}.

\section{Controllability and Observability}
\label{CandO}

The system of two interacting spin 
$1$'s,  if the gyromagnetic ratios 
are equal,  has dynamical Lie algebra 
isomorphic to $su(2)$ or $u(2)$ 
according to whether or not $J_{12}$ is equal to zero. 
In the case $J_{12}=0$ we also have a Lie algebra isomorphic to
$su(2)$ even in the case of different $\gamma$'s. 
The only nontrivial case is when $\gamma_1 \not= \gamma_2$ 
and $J_{12} \not=0$. In this
case, we have the following Theorem.  

\bt{CoTheo} If $\gamma_1 \not= \gamma_2$ 
and $J_{12} \not=0$, the system is  controllable namely the
dynamical Lie algebra is equal to su(9). 
\et \bpr 
 We have to prove that,  by calculating (repeated) Lie brackets
of the matrices $A$, $B_{x,y,z}$,  we can obtain all the matrices of the
form $iC \otimes D$ where $C$ and $D$ vary in the orthogonal basis of
$u(3)$ described  in the previous section,  
except the $9 \times 9 $ identity.  By repeated Lie brackets of the 
$B_{x,y,z}$ and using a determinant of Vandermonde type of argument
similar to the one in Lemma 4.1 of \cite{confraLAA}, 
 we  obtain all the matrices of the form $i\sigma \otimes {\bf 1}$
and     $i {\bf 1} \otimes \sigma $.  Then,  using  the Lie bracket of
these matrices with $A$ several
times, we  obtain also all the matrices of the form $i\sigma_1 \otimes
\sigma_2$. To obtain the other elements we proceed as follows: 
We calculate $[i \sigma_z \otimes \sigma_x, i\sigma_z \otimes \sigma_y]$
(see Table (\ref{Tavolaanticom})). This gives a
multiple of $i{\bf1} \otimes \sigma_z$ (which is already in the
dynamical Lie algebra) plus a multiple of $T \otimes \sigma_z$, with
$T$ defined in (\ref{Co4}). From this, taking the Lie brackets with
elements of the type $i {\bf 1} \otimes \sigma$ and 
$i {\sigma} \otimes  {\bf 1} $, using (\ref{plplp}) we obtain all the
matrices of the form $iS \otimes \sigma$ and analogously we can obtain 
all the matrices of the type $i \sigma \otimes S$. To 
conclude the proof of 
controllability we only have to prove that we can obtain all the
matrices $iS \otimes S$ except the $9 \times 9$ identity. Notice that
since $su(3)$ is a simple Lie algebra $[su(3),su(3)]=su(3)$. Therefore
given $C$ in $su(3)$  we can choose two matrices $M$
and $N$ such that $[M,N]=C$. Using the well known fact 
(see e.g. \cite{sakurai}) that $\sum_{j=x,y,z} \bar (-i\bar
\sigma_j)^2=2 \times {\bf
1}$, we   calculate 
\be{S14}
\sum_{j=x,y,z}[M\otimes -i\bar \sigma_{j}, N \otimes -i \bar 
\sigma_{j}]=\sum_{j=x,y,z}
 C\otimes (-i \bar \sigma_{j})^2=2C \otimes {\bf 1}. 
\ee
Analogously we can see that we can generate all the matrices ${\bf 1}
\otimes C$ with $C \in su(3)$. Now, since $C$ is a general 
matrix in $su(3)$ we can obtain all the
elements of the type $K \otimes Y$ with $Y \in su(3)$ (or $K \in su(3)$)
and $K$ (or $Y$) in the orbit $\oplus_{k=0}^\infty ad_{su(3)}^k i{\cal
S}$. However this orbit is equal to $su(3)$ (it is a nonzero ideal in
$su(3)$ and therefore it must be $su(3)$ itself since $su(3)$ is simple).
 This concludes the proof. 
\epr

\vs

In the case $\gamma_1 \not=\gamma_2$ $J_{12}\not=0$ the system being
controllable is also {\it observable}. In all the other cases, the
space $\cal V$ defined in (\ref{condob}) 
is different from $su(9)$. In these cases, initial density 
matrices which differ by a matrix in ${\cal V}^\perp$
cannot be distinguished and the system is not observable.

\section{Parameter Identification} 
\label{ParId}

We now characterize the classes 
of equivalent pairs model-initial 
state. In other terms, we investigate 
what can be said concerning the parameters 
of the system by experiments involving 
control with an external electro-magnetic field
and measurement of the total 
magnetization. We shall assume 
that we are in the controllable (and therefore observable) case,
namely we know that $\gamma_1 \not= \gamma_2$ and $J_{12}\not=0$.  
We state and prove the main result of 
this section in the following Theorem \ref{MainId}, where we characterize
the classes of equivalent models. In the following we mark with a prime
$'$ every symbol concerning system $\Sigma'$. 
We first give   three  preliminary results that  
can be proved  as in the case of  networks of spin
$\frac{1}{2}$'s treated in \cite{Spin12}. For completeness we give 
self contained proofs and some additional considerations 
in the Appendix.   
\bl{lemma1}
If, for every trajectory of $\Sigma$, $\rho$,
and corresponding trajectory of $\Sigma'$, $\rho'$,  we have
\be{C1}
Tr(S\rho)=Tr(S'\rho'), \qquad v=x,y,z, 
\ee
for some pair of matrices $S$ and $S'$, 
then for every $F$, $F:=ad_{B_{j_1}}ad_{B_{j_2}}\cdot \cdot \cdot
ad_{B_{j_r}}S$, and corresponding $F'$,
$F':=ad_{B'_{j_1}}ad_{B'_{j_2}}\cdot \cdot \cdot
ad_{B'_{j_r}}S'$, $(j_1,...,j_r \in \{x,y,z\}$ or $B_{j}=A$),
we have
\be{C2}
Tr(F\rho)=Tr(F'\rho'),
\ee
for every pair of trajectories $\rho$ and $\rho'$.
\el

\bl{lemma2}
Let  $(\Sigma, \rho_0)$  and $(\Sigma',\rho_0')$ be two equivalent
models. Then up to a permutation of the indices 
\be{gammas}
\gamma_{1,2}=\gamma_{1,2}', 
\ee
and for every $\sigma \in i{\cal S}$ 
\be{starting}
Tr(\sigma \otimes {\bf 1} \rho(t))=Tr(\sigma \otimes {\bf 1}\rho'(t)),
\qquad 
Tr({\bf 1}  \otimes \sigma  \rho(t))=Tr({\bf 1}  \otimes \sigma
\rho'(t)). 
\ee
\el 

\bl{lemma3}
Assume two models $(\Sigma,\rho_0)$ and $(\Sigma',\rho_0')$ are
equivalent.  
For every pair of matrices $S$ and $S'$ such that 
\be{mesmo}
Tr(S\rho)=Tr(S'\rho'), 
\ee 
we also have 
\be{m1}
Tr([S, \sigma \otimes {\bf 1}] \rho)=Tr([S', \sigma \otimes {\bf 1}]
\rho')
\quad 
Tr([S, {\bf 1} \otimes \sigma] \rho)=Tr([S', {\bf 1} \otimes \sigma] \rho').
\ee
\el

\vs 

We define now two orthogonal 
subspaces of $isu(9)$: $\cal I$ which is spanned by elements of the type
 $\sigma_1 \otimes \sigma_2$ and $S_1 \otimes S_2$ (namely the factors
of the tensor product are both in $i\cal S$ or both in $i{\cal
S}^\perp$),  except the identity,  and
${\cal I}^\perp$ which is spanned by 
mixed type of elements namely
elements of the type $\sigma \otimes S$ 
and $S \otimes \sigma$. We shall
use this decomposition of $isu(9)$ 
(which induces a decomposition of
$su(9)$) in the following treatment. The induced decomposition of
$su(9)$ is a Cartan type (see e.g. \cite{Helgason}) of decomposition as
stated in the following Lemma. 

\bl{commutazione}
\be{S28u}
su(9)=i{\cal I} \oplus i {\cal I}^\perp, 
\ee 
with 
\be{Cartan1}
[i{\cal I}^\perp, i{\cal I}^\perp] \subseteq  i{\cal I}^\perp, 
\ee
\be{Cartan2}
[i{\cal I}^\perp, i{\cal I}] \subseteq  i{\cal I}, 
\ee
\be{Cartan3}
[i{\cal I}, i{\cal I}] \subseteq  i{\cal I}^\perp.  
\ee
\el
\bpr
To verify (\ref{Cartan1}), we consider a Lie bracket $[\sigma_1 \otimes
S_1, \sigma_2 \otimes S_2]$ and prove that it is orthogonal to elements
of the form $\sigma_3 \otimes \sigma_4$ as well as to elements of the form
$S_3 \otimes S_4$. To do this, we rewrite  $[\sigma_1 \otimes
S_1, \sigma_2 \otimes S_2]$ as 
\be{rew}
[\sigma_1 \otimes
S_1, \sigma_2 \otimes S_2]=\sigma_1 \sigma_2 \otimes S_1 S_2 - 
\sigma_2 \sigma_1 \otimes S_2 S_1.  
\ee
We can decompose $\sigma_1 \sigma_2 \otimes S_1 S_2$ as 
\be{rew2}
\sigma_1 \sigma_2 \otimes S_1 S_2=\frac{1}{4}([\sigma_1,
\sigma_2]+\{\sigma_1, \sigma_2\} ) \otimes ([S_1,
S_2] + \{S_1,S_2\}). 
\ee
From this expression,  using (\ref{Co9}) and (\ref{nice}),  the
only term that is not perpendicular to $\sigma_3 \otimes \sigma_4$ is 
$ [\sigma_1,\sigma_2] \otimes [S_1,
S_2].$ Doing the same thing for  the second term on the right hand side of
(\ref{rew}), one obtains that the only term which is not perpendicular
 to $\sigma_3 \otimes \sigma_4$ is $[\sigma_2,\sigma_1] \otimes [S_2,
S_1].$ But these two terms cancel. Analogously one proves orthogonality
to matrices  of the type $S_3 \otimes S_4$. To conclude the proof of
(\ref{Cartan1}) one has to prove orthogonality of terms of the form 
 $[\sigma_1 \otimes
S_1, S_2 \otimes \sigma_2]$.  This is obtained using similar
arguments. Also similar arguments, considering all the 
sub-cases, prove (\ref{Cartan2}) and (\ref{Cartan3}). 
\epr

\br{estensione}
The argument in the above Lemma can be generalized to deal with
decompositions of $su(3^n)$, for every $n \geq 1$. 
One can define a subspace of $isu(3^n)$ of tensor products of matrices
of the form $\sigma \otimes S \otimes \cdot \cdot \cdot \otimes \sigma$
with an {\it odd} number of $\sigma$'s and a complementary space with an
even number of $\sigma$'s. Call these subspaces ${\cal I}_o$ and ${\cal
I}_e$ respectively. Then one can show, by induction on $n$ that 
\be{commugen}
[i{\cal I}_o, i{\cal I}_o] \subseteq i{\cal I}_o, \quad 
[i{\cal I}_o, i{\cal I}_e] \subseteq i{\cal I}_e,\quad 
[i{\cal I}_e, i{\cal I}_e] \subseteq i{\cal I}_o,
\ee  
and
\be{acommugen}
\{{\cal I}_o, {\cal I}_o\} \subseteq {\cal I}_e, \quad 
\{ {\cal I}_o, {\cal I}_e\} \subseteq {\cal I}_o,\quad 
\{ {\cal I}_e, {\cal I}_e \} \subseteq {\cal I}_e. 
\ee  
In fact for $n=1$ (\ref{commugen}) and (\ref{acommugen}) follow
immediately 
from (\ref{Co9}) and (\ref{nice}). For $n>1$, (\ref{commugen}) follows by
writing 
\be{identita}
[A  \otimes B, C \otimes D]=\frac{1}{2}( \{A,C\} \otimes [B,D]+[A,C]
\otimes \{ B,D \}), 
\ee 
and applying the inductive assumption to all the factors in this
expression and considering all the sub-cases. Analogously one can prove
(\ref{acommugen}). 
\er

\vs

The following is the main theorem of this section. 

\bt{MainId}
Two controllable pairs Model-Initial State 
 $\Sigma(n,J_{12},\gamma_1,\gamma_2,\rho_0)$,
$\Sigma'(n',J_{12}',\gamma_1',\gamma_2',\rho_0')$ are equivalent if and
only if (up to a permutation of the indices)
\begin{enumerate}
\item $\gamma_1=\gamma_1'$
\item $\gamma_2=\gamma_2'$
\item $|J_{12}|=|J_{12}'|$
\item
If $J_{12}=J_{12}'$ then $\rho_0=\rho_0'$. If $J_{12}=-J_{12}'$, denote
by 
$\rho_1$ ($\rho_1'$) the component of $\rho$ ($\rho'$)  in ${\cal
I}^\perp$
 and 
$\rho_2$ ($\rho_2'$) the component of $\rho$ ($\rho'$) in ${\cal I}$
then $\rho_1(0)=\rho_1'(0)$ and $\rho_2(0)=-\rho_2'(0)$.

\end{enumerate}
\et 
\bpr Assume first that the two pairs are equivalent. We have, 
from Lemma \ref{lemma2}, that, up to a permutation of
the indices, $\gamma_{1,2}=\gamma_{1,2}'$. 

Consider now the following procedure to generate a basis of
$su(9)$. Start with $A$, $i {\bf 1} \otimes \sigma$ and $i \sigma
\otimes {\bf 1}$ at Step $0$. At step $n$ take the Lie brackets of
the matrices obtained at step $n-1$ with $A$, 
$i {\bf 1} \otimes \sigma$ and $i \sigma
\otimes {\bf 1}$. By controllability, the 
procedure generates a basis of $su(9)$. 
Moreover every element we calculate belongs  to either $i{\cal I}$ or 
 $i{\cal I}^\perp$ and there are no combinations. This follows by
induction on the step and applying Lemma \ref{commutazione}. We can repeat the
same procedure starting with $A'$, $i {\bf 1} \otimes \sigma$ and $i \sigma
\otimes {\bf 1}$. Let $F$ and $F'$  be two corresponding matrices
obtained at a step $d\geq 1$. We
have  $F=J_{12}^k\bar F$ and $F'=J_{12}'^k \bar F$ for the same  $\bar
F$ and with $k$ odd for $F(F') \in i{\cal I}$ and even (not zero) for  
$F(F') \in i{\cal I}^\perp$. This is true at Step $1$ and follows by
induction for elements obtained at the following steps by applying Lemma
\ref{commutazione}. Now  notice that elements obtained from Step 1
on also span all of $su(9)$ as well as the elements obtained including
Step $0$. This is because,  if we call $R_1$ the subspace spanned by
elements obtained at Step 1, the elements from Step 1 on span the vector
space $\oplus_{k_1+k_2+\cdot \cdot \cdot +k_r=0}^\infty ad_{T_1}^{k_1}
ad_{T_2}^{k_2}\cdot \cdot \cdot ad_{T_r}^{k_r} R_1$, where $T_1,
T_2,...,T_r$ are in the set 
$\{ A, i{\bf 1}\otimes \sigma, i \sigma \otimes
{\bf 1}\}$. It follows from an application of the Jacobi identity 
that this is equal to $\oplus_{k=0}^\infty ad^k_{\cal
L}R_1$, where $\cal L$ is the Lie algebra generated by 
$\{ A, i{\bf 1}\otimes \sigma, i \sigma \otimes
{\bf 1}\}$ which by controllability is $su(9)$. So this is equal to  
$\oplus_{k=0}^\infty ad_{su(9)}^kR_1$ which is a nonzero ideal in $su(9)$
and therefore $su(9)$ itself since $su(9)$ is a simple Lie algebra. The
same argument holds with $A'$ replacing $A$. In conclusion we have, by
applying Lemmas \ref{lemma1}, \ref{lemma2} and \ref{lemma3}, {\it
for any} $\bar F \in {\cal I}$
\be{bas1}
J_{12}^kTr(\bar F \rho)=J_{12}'^kTr(\bar F \rho'), 
\ee 
with $k$ odd and, {\it for any} $\bar F \in {\cal I}^\perp$,
\be{bas2}
J_{12}^kTr(\bar F \rho)=J_{12}'^kTr(\bar F \rho'), 
\ee 
with $k$ even. In particular, by applying (\ref{bas2}) for $\bar F={\bf
1} 
\otimes \sigma$ and comparing with (\ref{starting}) of Lemma \ref{lemma2}
we obtain $|J_{12}|=|J_{12}'|$.

\vs 
The proof goes now in an analogous way to the case of spin $\frac{1}{2}$
treated in \cite{Spin12}. We have two cases: If $J_{12}=J_{12}'$, we
have $A=A'$, $B_{x,y,z}=B_{x,y,z}'$. In this case since the  systems  
are observable and we have the same input-output behavior then we must
have $\rho_0=\rho_0'$. If $J_{12}=-J_{12}'$,  then from the above
discussion we have 
\be{tracce}
Tr(G\rho_0)=Tr(G\rho_0'), \qquad \forall G \in {\cal I}^\perp,  
\ee   
and 
\be{tracce1}
Tr(G\rho_0)=-Tr(G\rho_0'), \qquad \forall G \in {\cal I},  
\ee  
so the components of $\rho_0$ and $\rho_0'$ in ${\cal I}^\perp$ are
equal while the components in ${\cal I}$ are opposite. 

\vs 

To prove the converse of the theorem, the only nontrivial case is when 
$J_{12}=-J_{12}'$.  In this
case let us write the equation for $\rho$ as
\be{C20a}
\dot \rho=[A+B(t),\rho],
\ee
and the equation for $\rho'$ as
\be{C21a}
\dot \rho'=[-A+B(t), \rho'].
\ee
We can write $\rho$ ($\rho'$) as $\rho:=\rho_1+\rho_2$,
$\rho':=\rho_1'+\rho_2'$  with $\rho_1(') \in {\cal I}^\perp$ and
$\rho_2(') \in \cal I$. Using relations (\ref{Cartan1}),
(\ref{Cartan2}), (\ref{Cartan3}) of Lemma \ref{commutazione} and
noticing that $A\in i{\cal I}$ while $B(t) \in i{\cal I}^\perp$, for every
$t$, we can write the differential equations for $\rho_1$ and $\rho_2$
as 
\begin{eqnarray}
\dot \rho_1=[B(t),\rho_1]+[A,\rho_2] \label{C23a}\\
\dot  \rho_2=[A,\rho_1]+[B(t),\rho_2] \nonumber,
\end{eqnarray}
and the differential equation for $\rho_1'$ and $\rho_2'$ as 
\begin{eqnarray}
\dot \rho_1'=[B(t),\rho_1']+[A,\rho_2']  \label{C24a}\\
\dot  \rho_2'=[A,\rho_1']+[B(t),\rho_2'] \nonumber.
\end{eqnarray}
Combining these equations we obtain a differential equation for
$\rho_1-\rho_1'$ and $\rho_2+\rho_2'$. In particular, we have
\begin{eqnarray}
\dot \rho_1 -\dot \rho_1'=
[B(t),\rho_1-\rho_1']+2[A,\rho_2+\rho_2'] \label{C25a}\\
\dot  \rho_2+ \dot \rho_2' =
2[A,\rho_1-\rho_1']+[B(t),\rho_2+\rho_2'] \nonumber.
\end{eqnarray}
From equations (\ref{C25a}), 
it follows that if $\rho_1(0)=\rho_1'(0)$ and
 $\rho_2(0)=-\rho_2'(0)$,  then $\rho_1(t)=\rho_1'(t)$ and
 $\rho_2(t)=-\rho_2'(t)$, for every $t$, 
 and for every control $B(t)$. In
particular, since $Tr(S_v^{TOT}\rho)=Tr(S_v^{TOT}\rho_1)$ 
and $\rho_1 \equiv \rho_1'$, 
the two models are equivalent.
\epr


\section{Conclusions}

We have presented a control theoretic analysis of a system of two
coupled 
spin $1$'s.  In particular,  this concerns the controllability,
observability and identifiability properties of this model. A Cartan
decomposition of the Lie algebra $su(3)$ induces a decomposition of the
Lie algebra $su(9)$ which plays a fundamental role in the
control theoretic properties of this system. A similar situation was
found in \cite{Spin12} for general networks of spin $\frac{1}{2}$ and it
is likely to appear for other type of networks of spins not necessarily
equal to $\frac{1}{2}$ or $1$. We believe that the methods of
analysis developed in this paper can be generalized to include for
example different forms of the interaction, models where one or more
components of the input field are held constant, networks with more than
two spins (cfr. Remark \ref{estensione}).   
We have proved that if (and only if) 
the gyromagnetic ratios are different  and the coupling constant is not
zero 
the system is controllable and
observable. In this case, we have characterized the set of equivalent 
models that give the same input output behavior. Our results are
motivated by the problem of identifying the unknown parameters in 
molecular magnets  through experiments involving driving  the
system with an input field and measuring   the total magnetization.
 The analysis is also instrumental to the design of  controls which will be
considered in further research.

\section*{Appendix: Proofs of Lemmas 
\ref{lemma1}, \ref{lemma2}, \ref{lemma3}}

\vs 

\noindent {\it Proof of Lemma \ref{lemma1}}

\vs

\bpr
The proof can be obtained by induction on the depth of $F$ and $F'$
defined as the number of the operations $ad$ in its calculation. For
depth $0$ (\ref{C2}) is the same as (\ref{C1}). Now assuming that $F$
has depth $d-1$, we can write for every $\tau$ and $t$ (considering a
trajectory corresponding to controls identically zero from a certain
instant  on)
\be{additional}
Tr(F e^{At} \rho(\tau) e^{-A t})= Tr(F'e^{A't} \rho'(\tau) e^{-A' t}), 
\ee  
which taking the derivative with respect to $t$ at zero gives 
\be{additional2}
Tr([A,F]\rho(\tau))=Tr([A',F']\rho'(\tau)). 
\ee 
Analogously one can obtain (with appropriate constant control) 
\be{additional3}
Tr([A+B_{x,y,z},F]\rho(\tau))=Tr([A'+B_{x,y,z}',F']\rho'(\tau)),  
\ee
which combined with  (\ref{additional2}) gives 
\be{additional4}
Tr([B_{x,y,z},F]\rho(\tau))=Tr([B_{x,y,z}',F']\rho'(\tau)).   
\ee
\epr

\vs

Lemma (\ref{lemma1}) can be generalized as follows. For
$\Sigma$ and $\Sigma'$ we can construct a basis for the dynamical Lie
algebra  starting from $A,B_{x,y,z}$ or $A',B_{x,y,z}'$ 
and at each step calculating the Lie brackets of the elements obtained at
the previous step by  $A,B_{x,y,z}$ or $A',B_{x,y,z}'$. 
Consider the depth of the element of the basis as the number of
Lie brackets calculated. The generalization consists of noticing that
if (\ref{C1}) holds for some $S$ and $S'$ it also holds for $[L,S]$ and
$[L',S']$ where $L$ and $L'$ are elements of the basis of the 
dynamical Lie Algebra obtained the same way just replacing the   
$A,B_{x,y,z}$ with  $A',B_{x,y,z}'$. This is true for every element of
depth $0$ from Lemma \ref{lemma1}. Now assume it is true for elements
$L$ and $L'$ of depth $d-1$. From the Jacobi identity we have 
\be{Jaco}
Tr([[B,L],S]\rho)+Tr([[L,S],B]\rho)+Tr([[S,B],L]\rho)=
\ee
$$
Tr([[B',L'],S']\rho')+Tr([[L',S'],B']\rho')+Tr([[S',B'],L']\rho')=0, 
$$
for some $B$ in the set $A,B_{x,y,z}$ and corresponding $B'$. Now the
second terms of the two sides are equal by applying the inductive
assumption and Lemma \ref{lemma1}. The same thing is true for the third
term where we apply first Lemma \ref{lemma1} to obtain 
$Tr([S,B]\rho)=Tr([S',B']\rho')$ and then the inductive assumption on
$L$ (with $S$ replaced by $[S,B]$). Therefore the first terms are also
equal. This facilitates the proofs of  Lemmas \ref{lemma2} and
\ref{lemma3}.

\vs 

\noindent {\it Proof of Lemma \ref{lemma2}}

\vs 

\bpr
By performing (repeated) Lie brackets of $B_x$, $B_y$ and $B_z$, it is
possible to obtain all the matrices of the form $
\gamma_1^k i \sigma \otimes {\bf 1} + \gamma_2^k i {\bf 1} \otimes \sigma 
$, with $k=1,2,...$ (cfr. Lemma 4.1. in \cite{confraLAA}). The
corresponding matrices for $\Sigma'$ are $
{\gamma_1'}^k i \sigma \otimes {\bf 1} + {\gamma_2'}^k 
i {\bf 1} \otimes \sigma.  
$
Now,  starting from  
\be{srt}
Tr(S_v^{TOT}\rho)=Tr(S_v^{TOT}\rho'), 
\ee
and taking the Lie bracket with the matrices above obtained,  we have 
\be{srt1}
\gamma_1^kTr(\sigma \otimes {\bf 1} \rho)+
\gamma_2^kTr({\bf 1} \otimes \sigma  \rho)
=\gamma_1'^k Tr(\sigma \otimes {\bf 1} \rho')+
\gamma_2'^kTr({\bf 1} \otimes \sigma \rho'), \quad k=0,1,2,...
\ee
Since  $Tr(\sigma \otimes {\bf 1} \rho)$ 
is not zero for every trajectory $\rho$ 
(unless $\rho$ is a scalar matrix which is a case we
exclude), the only possibility for (\ref{srt1}) to be verified
is that the determinant 
\be{Vander}
D=\left|\matrix{1 &1 & 1 &1 \cr 
\gamma_1 &\gamma_2 & \gamma_1' & \gamma_2' \cr
\gamma_1^2 &\gamma_2^2 & \gamma_1'^2 & \gamma_2'^2 \cr
\gamma_1^3 &\gamma_2^3 & \gamma_1'^3 & \gamma_2'^3} \right|, 
\ee
is equal to zero. But this is a Vandermonde determinant, therefore we
need two of the $\gamma$'s and $\gamma'$'s to be equal. 
Up to a permutation we can choose $\gamma_1=\gamma_1'$. We can now use
the same Vandermone determinant type of argument starting from 
\be{Vander2}
\gamma_1^k(Tr(\sigma \otimes {\bf 1} \rho)- Tr(\sigma \otimes {\bf 1}
\rho'))+\gamma_2^kTr({\bf 1} \otimes \sigma  \rho)-
\gamma_2'^kTr({\bf 1} \otimes \sigma \rho')=0, 
\ee
to  conclude that $\gamma_2=\gamma_2'$ and that (\ref{starting}) holds. 
\epr

\vs 

\noindent {\it Proof of Lemma \ref{lemma3}}

\vs

\bpr
As in the previous Lemma we obtain 
\be{m2}
\gamma_1^k(Tr(\rho[S, \sigma \otimes {\bf 1}])-Tr(\rho'[S',\sigma
\otimes {\bf 1}])+\gamma_2^k(Tr(\rho[S, {\bf 1} \otimes \sigma])
-Tr(\rho'[S',{\bf 1} \otimes {\bf 1}])
=0, k=1,2,... 
\ee
This since $\gamma_1 \not= \gamma_2$, we obtain (\ref{m1}). 
\epr

\end{document}